\documentclass[12pt,sort&compress]{elsarticle}

\makeatletter                                   %
\def\NAT@def@citea{\def\@citea{\NAT@separator}} 
\makeatother                                    %

\usepackage[utf8]{inputenc}
\usepackage{amsmath}
\usepackage{amssymb}
\usepackage{epstopdf}

\usepackage{enumitem} 
\usepackage{pdflscape} 

\begin{document}

\renewcommand{\figurename}{Figure} 
\renewcommand{\tablename}{Table} 

\begin{frontmatter}

\title{Annual modulation of dark matter:\\The ANAIS--112 case
}

\author[mainaddress,secondaryaddress]{I. Coarasa}
\author[mainaddress,secondaryaddress]{J.~Amaré}
\author[mainaddress,secondaryaddress]{S.~Cebrián}
\author[mainaddress,secondaryaddress,tertiaryaddress]{C.~Cuesta}
\author[mainaddress,secondaryaddress]{E.~García}
\author[mainaddress,secondaryaddress,quaternaryaddress]{M.~Martínez}
\author[mainaddress,secondaryaddress]{M.A.~Oliván}
\author[mainaddress,secondaryaddress]{Y.~Ortigoza}
\author[mainaddress,secondaryaddress]{A.~Ortiz de Solórzano}
\author[mainaddress,secondaryaddress]{J.~Puimedón \fnref{contactperson}}
\author[mainaddress,secondaryaddress]{M.L.~Sarsa}
\author[mainaddress,secondaryaddress]{J.A.~Villar}
\author[mainaddress,secondaryaddress]{P.~Villar}

\fntext[contactperson]{\textit{Email address:} puimedon@unizar.es}

\address[mainaddress]{Grupo de Física Nuclear y Astropartículas, Universidad de Zaragoza, Calle Pedro Cerbuna 12, 50009 Zaragoza, Spain}
\address[secondaryaddress]{Laboratorio Subterráneo de Canfranc, Paseo de los Ayerbe s/n, 22880 Canfranc Estación, Huesca, Spain}
\address[tertiaryaddress]{\textit{Present Address:} Centro de Investigaciones Energéticas, Medioambientales y
Tecnológicas, CIEMAT, 28040, Madrid, SPAIN}
\address[quaternaryaddress]{\textit{Present Address:} Università di Roma La Sapienza, Piazzale Aldo Moro 5, 00185 Roma, Italy}

\begin{abstract}

The annual modulation measured by the DAMA/LIBRA experiment can be explained by the interaction of dark matter WIMPs in NaI(Tl) scintillator detectors. Other experiments, with different targets or techniques, exclude the region of parameters singled out by DAMA/LIBRA, but the comparison of their results relies on several hypotheses regarding the dark matter model. ANAIS--112 is a dark matter search with 112.5~kg of NaI(Tl) scintillators under commissioning at the Canfranc Underground Laboratory (LSC) to test the DAMA/LIBRA result in a model independent way. We analyze its prospects in terms of the \textit{a priori} critical and detection limits of the experiment. The analysis is based on the detector response and the background level measured for the first modules operated in Canfranc.

\end{abstract}


\end{frontmatter}


\section{Introduction}

The ANAIS project is intended to search for dark matter annual modulation with ultrapure NaI(Tl) scintillators at the Canfranc Underground Laboratory (LSC) in Spain, in order to provide a model independent confirmation of the signal reported by the DAMA/LIBRA collaboration \cite{Bernabei20131} using the same target and technique \cite{kims2016_epjc,dmice2014_prd,sabre_taup2013,picolon_taup2015}. The WIMP interaction counting rate experiences an annual modulation
as the result of the motion of the Earth around the Sun that can be approximated \cite{Freese_modulation88,Savage20091} by:
\begin{equation}
\frac{dR}{dE_R}\left(E_R,t\right)\approx S_0\left(E_R\right)+S_m\left(E_R\right)\cdot{}\textnormal{cos}\left(2\pi\frac{t-t_0}{T}\right),
\label{eq:annualModulation}
\end{equation}
where $R$ is the interaction rate, $E_R$ is the recoil energy, $t_0$ is the expected time of the maximum (or minimum, depending on the sign of $S_m$), about 150 days after 1st January, and $T$ is the expected period of one year. The time-averaged differential rate is denoted by $S_0$, whereas the modulation amplitude is given by $S_m$ \cite{Savage20091}. The measured value of $S_m$ by DAMA/LIBRA is $0.0112\pm0.0012$ cpd/kg/keV${_\textnormal{ee}}$ within [2,6] keV$_{\textnormal{ee}}$ interval (cpd stands for \textit{counts per day} and keV${_\textnormal{ee}}$ for keV electron--equivalent). In this paper we will show that ANAIS--112 can observe this modulation amplitude.

ANAIS--112 consists of nine modules of 12.5~kg each, made by Alpha Spectra, Inc. Colorado with ultrapure NaI powder.
The so called D0, D1 and D2 modules took data in ANAIS--25 (D0 and D1) \cite{anais25} and ANAIS--37 (D0, D1 and D2) \cite{anais37} set--ups. We estimated their cosmogenic activation \cite{jcap_cosmo_d0d1}, measured their background after the decay of the short--lived cosmogenic radioisotopes and elaborated a satisfactory background model \cite{epjc_bkg_d012}. The three modules have an excellent light yield of 15 photoelectrons/keV \cite{anais37,light_yield}.
Three new modules (D3, D4 and D5) were received along 2016 and tested in ANAIS--37 set--up.
Regarding crystals radiopurity, $^{40}$K and $^{210}$Pb dominate the contribution to the low energy region. Table \ref{table:40K-210Pb} shows the measured specific activity for D0 to D5.
The last three modules (D6, D7 and D8) arrived at LSC in March 2017 and are being characterized. We expect a $^{40}$K and $^{210}$Pb content similar to that of D5 because the four modules has been made from the same ingot.

The $3\times 3$ configuration of ANAIS--112 is currently installed at the hall B of LSC under 2450 m.w.e. It consists of a shielding of 10 cm of old lead, 20 cm of low activity lead, 40 cm of a neutron moderator and an anti--radon box. An active muon veto system covers the top and sides of the set--up.

\section{Searching a signal in the [2,6] keV$_{\textnormal{ee}}$ energy interval} \label{sec:2to6}

A model independent way to check the DAMA/LIBRA result is looking for a signal not only with the same target and technique but also in the same region where DAMA/LIBRA finds it.
As the signal could be of unknown origin not related with dark matter, we will evaluate in section \ref{ssec:2to6_noDM_hyp} the detection in [2,6] keV$_{\textnormal{ee}}$ of an annual modulation amplitude $b$ of the counting rate $B$
\begin{equation}
B(\tau)=a+b\textnormal{ cos }\tau,
\label{eq:uniformBkg}
\end{equation}
where $a$ is the mean annual rate and $\tau=2\pi(t-t_0)/T$, see Eq.~(\ref{eq:annualModulation}).
We will consider the simplest approximation of only one 4~keV$_{\textnormal{ee}}$ bin; afterwards we will take into account the energy binning and the segmentation of the 112.5~kg in 9 modules.
In section \ref{ssec:2to6_DM_hyp}, we will consider the particular case of a modulation induced by dark matter \cite{Cebrian2001339}.

\subsection{Model independent modulation} \label{ssec:2to6_noDM_hyp}

The test statistic \cite{Eadie_1971} to evaluate the null ($b=0$) and the alternative ($b\ne0$) hypotheses is the least squares estimator of the amplitude, $\hat{b}$, of expected value $E(\hat{b})=b$ and variance $var(\hat{b})$. Asymptotically, $\hat{b}$ follows a normal distribution.

The critical limit ($L_C$) is a threshold such that if $\hat{b}> L_C$, the signal is statistically significant. $L_C$ is defined from the distribution of $\hat{b}$ when there is no signal, $E(\hat{b})=0$. We use a one-tailed test because the amplitude measured by DAMA/LIBRA is positive. For a confidence level $\alpha$, the probability of a false positive is $1-\alpha$:
\begin{equation}
P(\hat{b}\leq L_C\mid b=0)=\alpha
\end{equation}

The detection limit ($L_D$) is the modulation amplitude such that the outcome of its estimator $\hat{b}$ is greater than $L_C$ with $\beta$ probability:
\begin{equation}
P(\hat{b}>L_C\mid b=L_D)=\beta
\end{equation}

\subsubsection{A single energy bin}

A linear least-squares fit \cite{Bevington200398} of Eq.~(\ref{eq:uniformBkg}) for $n$ time bins, where $B_i$ is the measured rate in the $i$th time bin $\tau_i$ and $w_i=1/var(B_i)$, gives:
\begin{equation}
\hat{b}=\frac{\sum_l{w_lB_l\cdot{}\left(-\sum_i{w_i\textnormal{cos }\tau_i}+\textnormal{cos }\tau_l\cdot{}\sum_i{w_i}\right)}}{\sum_i{w_i}\cdot{}\sum_i{w_i\textnormal{cos}^2\tau_i}-\left(\sum_i{w_i\textnormal{cos }\tau_i}\right)^2}
\end{equation}
\begin{equation}
var(\hat{b})=\frac{\sum_i{w_i}}{\sum_i{w_i}\cdot{}\sum_i{w_i\textnormal{cos}^2\tau_i}-\left(\sum_i{w_i\textnormal{cos }\tau_i}\right)^2}
\label{eq:estimator_Variance}
\end{equation}

We can obtain a simple expression for $var(\hat{b})$. If $N_i$ is the number of observed events (Poisson distributed) and $\varepsilon$ is the fraction of true events remaining after the cuts to reject the noise and select true events:
\begin{equation}
B_i=\frac{N_i/\varepsilon}{\Delta E\cdot{}M\cdot{}\Delta t} \hspace{1.0 cm} var(B_i)=\frac{B_i/\varepsilon}{\Delta E\cdot{}M\cdot{}\Delta t}
\end{equation}
where $M$ is the total detection mass, $\Delta E$ and $\Delta t$ are the width of the energy and time bins, respectively.

If $b=0$, the expected value $E(B_i)=B$ is time independent. For $b\neq0$ $E(B_i)$ is nearly time independent if $b\ll a$, as is the case for the annual modulation measured by DAMA/LIBRA ($b\sim10^{-2}$ cpd/kg/keV$_{\textnormal{ee}}$) and the typical counting rates ($a\gtrsim 1$ cpd/kg/keV$_{\textnormal{ee}}$). Latter value guarantees also the normality of $\hat{b}$ for ANAIS--112, even for one day time bins. Then:
\begin{equation}
var(B_i)\approx \frac{B/\varepsilon}{\Delta E\cdot{}M\cdot{}\Delta t}=\frac{1}{w}
\end{equation}
and if an integer number of periods is measured and the time bins are equally spaced, Eq.~(\ref{eq:estimator_Variance}) is simplified because
$\sum_{i}{w_i\textnormal{cos }\tau_i}\simeq w\sum_{i}{\textnormal{cos }\tau_i}=~0$ and $\sum_{i}{w_i\textnormal{cos}^2\tau_i}\simeq w\sum_{i}{\textnormal{cos}^2\tau_i}\simeq w\cdot{}n\cdot{}\frac{1}{2}$; then, $var(\hat{b})$ is
\begin{equation}
var(\hat{b})=\frac{2\cdot{}B}{\Delta E\cdot{}M\cdot{}T_M\cdot{}\varepsilon} \hspace{1.0cm} \left(b\ll a\right)
\label{eq:bVariance}
\end{equation}
with $T_M=n\cdot{}\Delta t$ the measurement time.

$L_C$ and $L_D$ are proportional to the standard deviation $\sigma(\hat{b})=\sqrt{var({\hat{b})}}$, which can be used as a figure of merit to compare the different experiments looking for the annual modulation observed by DAMA/LIBRA.
\begin{equation}
FOM=\left({\frac{2\cdot{}B}{\Delta E\cdot{}M\cdot{}T_M\cdot{}\varepsilon}}\right)^{\frac{1}{2}}
\label{eq:fom}
\end{equation}

\subsubsection{Background estimate}

We have to assess $B$ and $\varepsilon$ of the nine modules from the current data of the six modules D0 to D5. The background rates of modules D0, D1 and D2 have been measured and are well known \cite{epjc_bkg_d012}. The rates of D3, D4 and D5 were measured when the activities of the short--lived cosmogenic isotopes were appreciable. We have to estimate their expected rates after the decay of these isotopes.

The background of a module in [2,6] keV$_{\textnormal{ee}}$ can be written as
\begin{equation}
B=A(K)f(K)+A(Pb)f(Pb)+B^\prime
\end{equation}
where $A(K)$ is the specific activity of $^{40}$K inside the crystal, $f(K)$ is the conversion factor from mBq to cpd/keV$_{\textnormal{ee}}$ in a module in anticoincidence with the other modules and, similarly, $A(Pb)$ and $f(Pb)$ for $^{210}$Pb. $B^\prime$ is the rate from other sources, mainly the photomultiplier tubes (PMT) and the long-lived cosmogenic isotopes ($^{22}$Na and $^{3}$H) \cite{epjc_bkg_d012}.

The conversion factors for the ANAIS--37 set--up, where D2 was between D0 and D1, have been estimated by Monte Carlo \cite{epjc_bkg_d012}. The factor $f(K)$ of D2 is less than the one of D0 or D1 (Table \ref{table:detectorsBkg}) because to detect the K--binding energy of Ar (3.2--keV), the 1.46--MeV gamma emitted in the $^{40}$K electron capture must not interact in any module. The difference between $f(Pb)$ of D0 or D1 and D2 is due to the different proportion of $^{210}$Pb in the bulk and on the surface of the NaI(Tl) crystals \cite{epjc_bkg_d012}.
The values of $B^\prime$ for D0, D1 and D2 obtained from the estimated of $^{40}$K and $^{210}$Pb contributions and the measured background $B$, are listed in Table~\ref{table:detectorsBkg}. D0 and D1 are equivalent as they were produced from the same ingot, with identical protocols, and sent to LSC simultaneously. $B^\prime$ is slightly lower in D2 because a different ingot was used, the manufacturing process was improved and it was exposed to cosmic rays for a shorter time span \cite{epjc_bkg_d012}.

The expected background rate $B$, after the decay of the short--lived cosmogenic isotopes, of modules D3 to D5 can be estimated assigning to them the means of $B^\prime$, $f(K)$ and $f(Pb)$ from the two data sets we have (the first one for the couple D0--D1, with $B^\prime$=2.35$\pm$0.16~cpd/kg/keV$_{\textnormal{ee}}$, and the second one for D2).
Their standard deviations, that reflect the different conditions of the modules D0, D1 and D2, can be regarded as systematic uncertainties.
Finally, we suppose that D6, D7 and D8 contain the same specific activity of $^{40}K$ and $^{210}Pb$ than D5 because they have been made from the same ingot and with the same production process.
The results are listed in the Table \ref{table:detectorsBkg}. We have taken the factors $f(K)$ of ANAIS--112 similar to those of ANAIS--37 though they are geometry dependent. Since in ANAIS--112 they are smaller because each module has more modules around, our estimate of $B$ is conservative.

The cut efficiencies, $\varepsilon$, have been estimated for D0, D1, D2 and D3 modules \cite{thesis_patr}. They are comparable in the [2,6] keV$_{\textnormal{ee}}$ interval, with average $\varepsilon=0.89$.

\subsubsection{Critical limit and detection limit}\label{sssec:LC_LD}

The usual results of the experiments looking for dark matter are exclusion plots (upper limits) at 90\% C.L. in the plane cross section WIMP--nucleon versus WIMP mass \cite{Savage20091}.
By definition of $L_D$, any upper limit, $L_U$, satisfies $L_U\leq L_D$, if both are set to the same C.L. and $var(\hat{b})\simeq var(\hat{b}\mid b=0)$ \cite{currie_1968}. ANAIS--112 fulfills the latter condition because $b\ll a$, see Eq.~(\ref{eq:bVariance}).
Furthermore, $L_C\leq L_U$ (both to the same C.L.) if the outcome of $\hat{b}\geq0$. If $\hat{b}<0$, it should not be very negative because if $\hat{b}\ll-\sigma(\hat{b})$, it would imply a negative modulation, opposite to the oberved by DAMA/LIBRA. Briefly, any $L_U$ given by ANAIS--112 will be less than $L_D$, likely greater than $L_C$ or, at least, not much smaller than $L_C$.

Therefore, in order to compare properly the expectations of ANAIS--112 with other experiments, we also chose the 90\% C.L. for $L_C$ and $L_D$. Then, $L_C=1.28~\sigma(\hat{b})$ and $L_D=2L_C$. Using Eq.~(\ref{eq:bVariance}) with $B=3.93$~cpd/kg/keV$_{\textnormal{ee}}$ (Table \ref{table:detectorsBkg}), $\Delta E=4$ keV$_{\textnormal{ee}}$, $M=112.5$ kg, $T_M=5$ years, $\varepsilon=0.89$ and taking into account the uncertainty of the estimated background (Table \ref{table:detectorsBkg}):
\begin{equation}
L_D=(8.40\pm 0.25)\cdot 10^{-3} \textnormal{ cpd/kg/keV}_{\textnormal {ee}} \hspace{0.5cm}\left(90\%\textnormal{ C.L.}\right)
\label{eq:aSingleEnergyBin}
\end{equation}
that is less than the DAMA/LIBRA signal. Then, ANAIS--112 can detect it. Furthermore, if the estimator of the DAMA/LIBRA signal is normal, with mean and standard deviation 0.0112 and 0.0012 cpd/kg/keV${_\textnormal{ee}}$, respectively, less than 1\% of the probability distribution is below our central value for $L_D$.

It is worth citing that, assuming a background linearly decreasing with time as an approximation of the decay of the long--lived $^{210}$Pb and $^{3}$H during data taking, the obtained $L_D$ is very similar to the one obtained assuming a constant background \cite{TFM_ivan}.

\subsubsection{Energy binning and segmented detector} \label{sssec:2to6_binned}

The energy binning and the segmented detector in nine modules should be considered to get more accurate $L_C$ and $L_D$ and to obtain the possible energy dependence of the modulation amplitude $b(E)$.

We estimate the spectra of the D3 to D8 modules scaling the D2 spectrum to get their respective backgrounds in [2,6] keV$_{\textnormal{ee}}$ (Table \ref{table:detectorsBkg}).
This can be justified because, though the contamination of D0 (or D1) is different to that of D2, the spectrum of D0 (or D1) is nearly proportional to the D2 one in [2,6] keV$_{\textnormal{ee}}$ \cite{Amare20161}.
For [1,6] keV$_{\textnormal{ee}}$ (section \ref{sec:1to6}), we conserve the former result, modifying only the [1,2] keV$_{\textnormal{ee}}$ to get the backgrounds of Table \ref{table:detectorsBkg[1,6]}.

\begin{figure}
	\centering
		\includegraphics[height=7cm]{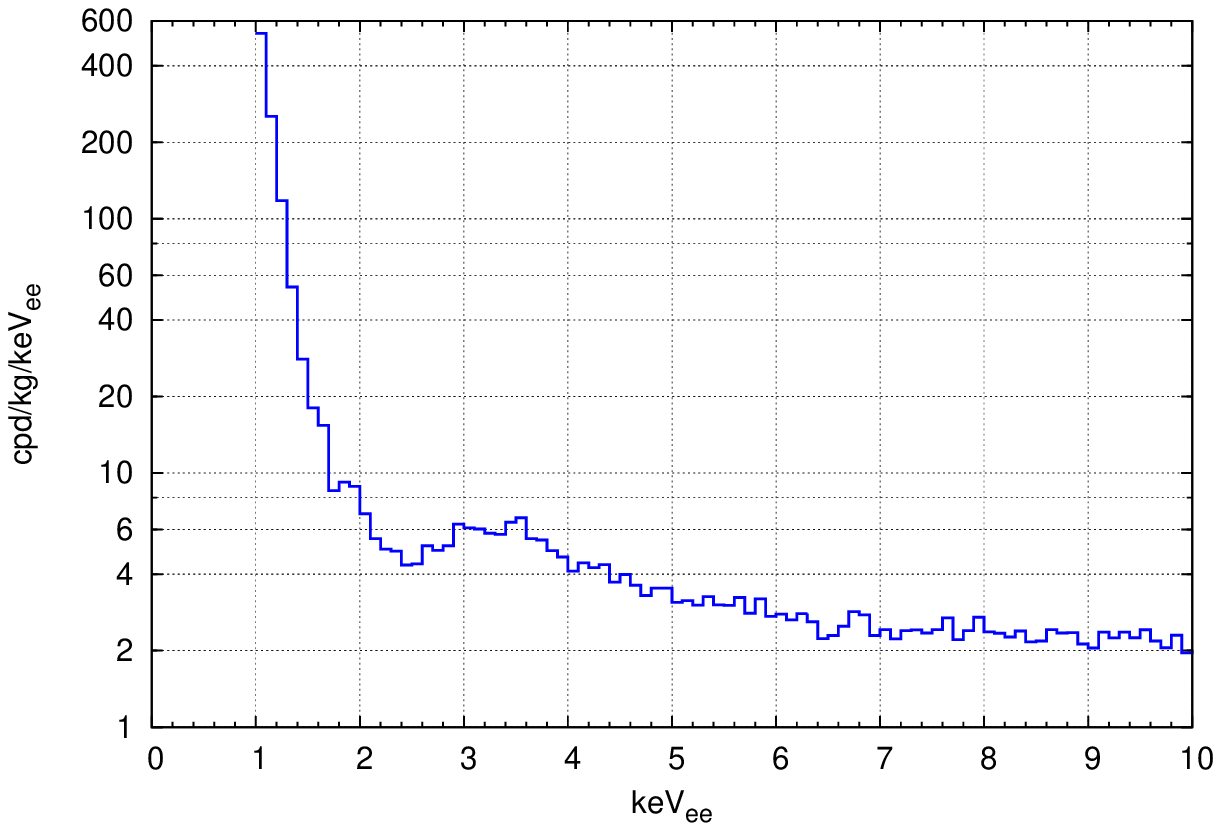}
	\caption{Estimated average background of ANAIS--112 at low energy corrected for efficiencies. This spectrum is valid only within [1,6] keV$_{\textnormal{ee}}$ (see section \ref{sssec:2to6_binned}).\label{fig:background}}
\end{figure}

The efficiencies $\varepsilon(E)$ are very similar for the measured D0 to D3 modules \cite{thesis_patr}. That of D2 is an intermediate function, which has be assumed valid for the remaining five modules, D4 to D8. The ANAIS--112 estimated background at low energy, valid for [1,6] keV$_{\textnormal{ee}}$ interval, and corrected by efficiencies is shown in Fig.~\ref{fig:background}.

\begin{enumerate}[label=($\alph*$)]
	\item \textit{Energy binning}
\end{enumerate}

The average annual modulation amplitude in the [2,6] keV$_{\textnormal{ee}}$ interval is
\begin{equation}
b=\frac{1}{\Delta E}\int_{E_1}^{E_1+\Delta E}b\left(E\right)dE,
\label{eq:bAmplitude}
\end{equation}
where $E_1=2$ and $\Delta E=4$~keV$_{\textnormal{ee}}$. Then, for $N$ bins the $j$th modulation amplitude in $\left[E_j,E_{j+1}\right]$ ($j=1,2,\cdots,N$) is
\begin{equation}
b_j=\frac{1}{\Delta E_j}\int_{E_j}^{E_{j+1}}b\left(E\right)dE,
\label{eq:bjAmplitude}
\end{equation}
being $\Delta E_j=E_{j+1}-E_{j}$. According to Eq.~(\ref{eq:bVariance})
\begin{equation}
var(\hat{b}_j)=\frac{2\cdot{}B_j}{\Delta E_j\cdot{}M\cdot{}T_M\cdot{}\varepsilon_j}
\label{eq:bjVariance}
\end{equation}
where $B_j$ and $\varepsilon_j$ are the background and the efficiency in the $j$th bin, respectively.
If all the bins are of equal width $\Delta E_j=\Delta E/N\equiv \delta E$, then
\begin{equation}
b=\frac{1}{N\cdot{}\delta E}\sum_{j=1}^{N}{\int_{E_j}^{E_{j+1}}b\left(E\right)dE}=\frac{1}{N}\cdot{}\sum_{j=1}^{N}{b_j}
\end{equation}
so that $b$ is the arithmetic mean of $b_j$.
For $N=40$ ($\delta E=0.1$~keV$_\textnormal{ee}$), a rate of 4~cpd/kg/keV$_\textnormal{ee}$ and $M=112.5$~kg, $\hat{b}_j$'s are virtually normal variables for one day time bins. When the $\hat{b}_j$'s are statistically independent
\begin{equation}
var(\hat{b})=\frac{1}{N^2}\cdot{}\sum_{j=1}^{N}var(\hat{b}_j)=\frac{2\cdot{}\left\langle B/\varepsilon \right\rangle}{\Delta E\cdot{}M\cdot{}T_M}
\label{eq:bVarianceBinning}
\end{equation}
with $\left\langle B/\varepsilon \right\rangle=(1/N)\cdot{}\sum_{j=1}^{N}{B_j/\varepsilon_j}$, see last row of Table \ref{table:detectorsBkg}.

\begin{enumerate}[resume, label=($\alph*$)]
	\item \textit{Segmented detector}
\end{enumerate}

We consider now the data of each module. According to Eq.~(\ref{eq:bjVariance}), the variance of the estimator of the modulation amplitude in the $j$th energy bin of the module $k$ ($k=1,2,\cdots,9$) is:
\begin{equation}
var(\hat{b}_j^k)=\frac{2\cdot{}B_j^k}{\delta E\cdot{}m\cdot{}T_M\cdot{}\varepsilon_j^k}
\end{equation}
where $m=12.5$ kg is the mass of one module and $B_j^k$ and $\varepsilon_j^k$ are the background and the efficiency in the $j$th energy bin of the module $k$. Now, $\hat{b}_j^k$'s are virtually normal variables for one week time bins. Thus, the variance of the estimator of $b$ in the module $k$ is:
\begin{equation}
var(\hat{b}^k)=\frac{2\cdot{}\left\langle B/\varepsilon \right\rangle^k}{\Delta E\cdot{}m\cdot{}T_M}
\end{equation}
with $\left\langle B/\varepsilon \right\rangle^k=(1/N)\cdot{}\sum_{j=1}^{N}{B_j^k/\varepsilon_j^k}$. The estimator $\hat{b}$ with the nine modules is the weighted mean of the nine $\hat{b}^k$ and its variance is:
\begin{equation}
var(\hat{b})=\left(\sum_{k=1}^{9}{\frac{1}{var(\hat{b}^k)}}\right)^{-1}=\frac{2}{\Delta E\cdot{}m\cdot{}T_M}\cdot{}\left(\sum_{k=1}^{9}{\frac{1}{\left\langle B/\varepsilon\right\rangle^k}}\right)^{-1}
\label{eq:9detectorsVariance}
\end{equation}

According to the Table \ref{table:detectorsBkg}, $L_D=8.27\cdot{}10^{-3}$ cpd/kg/keV$_\textnormal{ee}$, very close to Eq.~(\ref{eq:aSingleEnergyBin}) because the nine values ${\left\langle B/\varepsilon\right\rangle}^k$ are close to ${\left\langle B/\varepsilon\right\rangle}$ and $B(E)/\varepsilon(E)$ is smooth enough to be $\approx B/\varepsilon$.

\subsection{Dark matter hypothesis} \label{ssec:2to6_DM_hyp}

This hypothesis means that the possible modulation has to be compatible with the energy dependence of the modulation amplitude, $b(E;\sigma,M_W)$ \cite{primack_seckel_sadoulet}, where ~$\sigma$ is the WIMP--nucleon cross section and $M_W$ the WIMP mass.
We take the differential rate from \cite{Savage20091}, the local dark matter density $\rho=0.3$~GeV/cm$^3$, the most probable WIMP velocity $v_0=220$ km/s and the escape velocity $v_{esc}=544$ km/s \cite{Smith2007755}.
We consider the spin-independent WIMP interaction, using the Helm nuclear form factor \cite{Lewin199687} and $Q_{Na}=0.30$ and $Q_I=0.09$ for the sodium and iodine quenching factors to transform the nuclear recoil energy into electron equivalent one, respectively \cite{Savage20091}.
The resolution, that has been measured for the D0, D1 and D2 modules, can be approximated in [1,6] keV$_\textnormal{ee}$ by $\Gamma/E=1.19 [E(\textnormal{keV})]^{-1/2}-0.016$, where $\Gamma$ is the full width at half maximum \cite{light_yield}.
The Earth velocity \cite{Bernabei2013370} is given by
\begin{equation}
v_E(t)=232+15\textnormal{ cos}\left(2\pi\frac{t-152.5}{365.25}\right) \textnormal{ km/s,}
\label{eq:EarthVelocity}
\end{equation}
with the maximum value at $t=152.5$ days (2nd June). 

The test statistic in this case is the maximum likelihood ratio, which we already used in a more general context \cite{Cebrian2001339}. It is asymptotically equivalent to test the difference between the $\chi^2_{min}$ of the null ($\sigma=0$) and alternative ($\sigma\ne0$) hypotheses \cite{Eadie_1971}. This equivalence is easily satisfied for 0.1~keV$_\textnormal{ee}$, see section \ref{sssec:2to6_binned}. The minimum of
\begin{equation}
\chi^2(\sigma,M_W)=\sum_{j}{\cfrac{\left(\hat{b}_j-b_j(\sigma,M_W)\right)^2}{var(\hat{b}_j)}},
\label{eq:chi2}
\end{equation}
has to be evaluated for $\sigma=0$ and $\sigma\ne0$. If $\sigma=0$, the quantity
\begin{equation}
\Delta\chi^2=\chi^2(\sigma=0,M_W)_{min}-\chi^2(\sigma,M_W)_{min}
\end{equation}
is distributed as a $\chi^2_\nu$ variable with $\nu=2$ degrees of freedom. $L_C$ at 90\% C.L. is such that $P(\chi^2_2\leq L_C)=0.9$, $L_C=4.61$.
On the other hand, if $\sigma\ne0$, $\Delta\chi^2$ is a non--central $\chi'^2_{(\nu,\lambda)}$ with $\nu=1$ degree of freedom, expected value
\begin{equation}
\left\langle\Delta\chi^2\right\rangle=\frac{1}{2}\cdot{}\sum_{j}{\cfrac{b_j^2\cdot{}\Delta E_j\cdot{}\varepsilon_j}{B_j}\cdot{}M\cdot{}T_M}+2,
\label{eq:likelihoodRatio}
\end{equation}
(see Ref. \cite {Cebrian2001339}) and non--central parameter $\lambda=\left\langle\Delta\chi^2\right\rangle-1$. The detection limit at 90\% C.L. is defined by $P(\chi'^2_{(1,\lambda)}>L_C)=0.9$, that holds when $\left\langle\Delta\chi^2\right\rangle=12.8$.

The segmented detector can be incorporated to the test, obtaining
\begin{equation}
\left\langle\Delta\chi^2\right\rangle=\frac{1}{2}\cdot{}\sum_{j,k}{\cfrac{(b_j^k)^2\cdot{}\Delta E_j\cdot{}\varepsilon_j^k}{B_j^k}\cdot{}m\cdot{}T_M}+2.
\label{eq:likelihoodRatioSpatialBinning}
\end{equation}

\begin{figure}
	\centering
		\includegraphics[height=7.5cm]{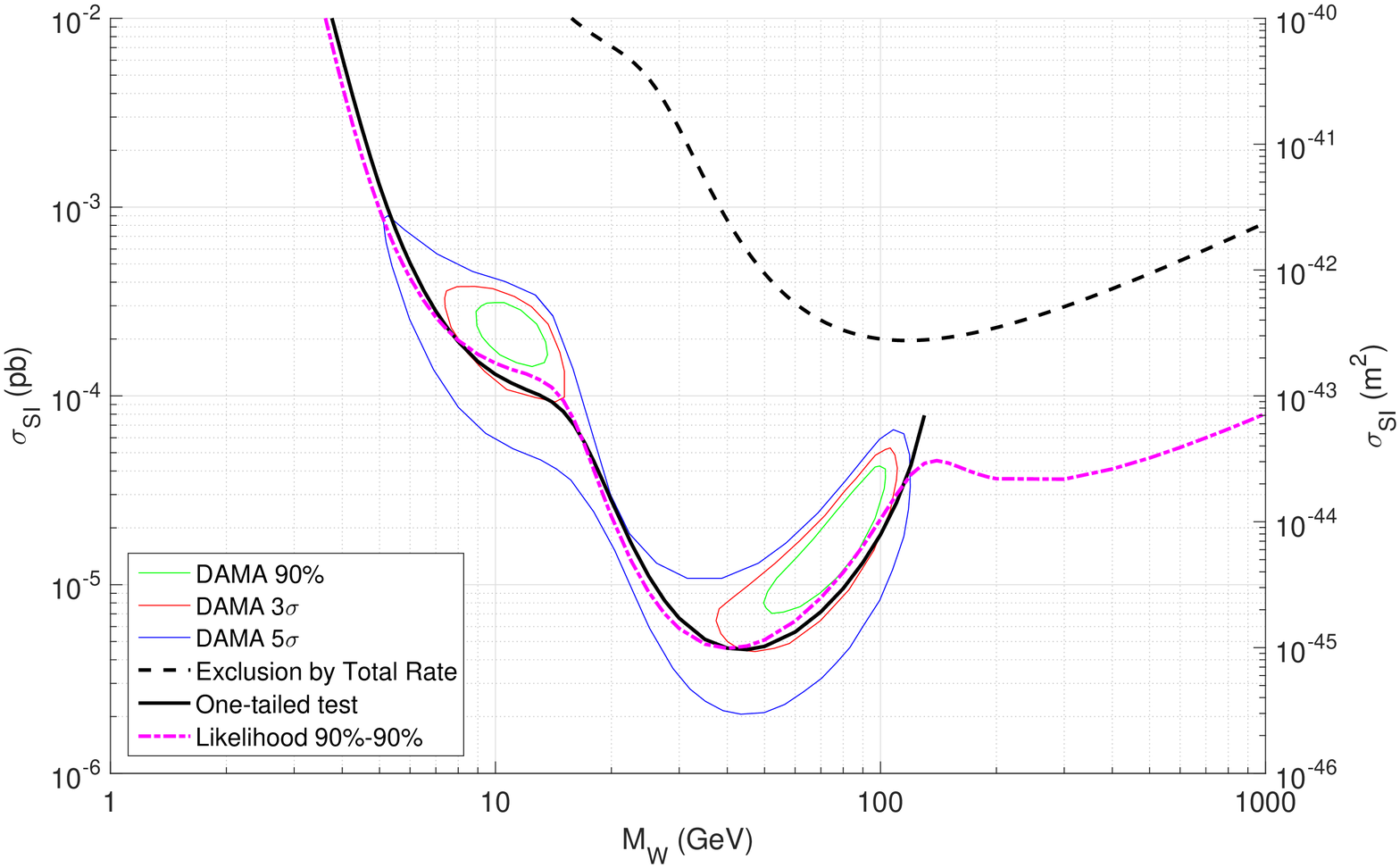}
	\caption{Result of the maximum likelihood test ratio for the detection limit in the [2,6] keV$_{\textnormal{ee}}$ window at 90\% C.L. (when critical limit is at 90\% C.L.) with 40 energy bins and segmented detector of ANAIS--112 after 5 years of measurement (dashed-dot magenta line). The exclusion by total rate for the spin-independent WIMP-nucleon cross section of ANAIS--112 is the dashed black line. DAMA/LIBRA regions at 90\% (solid green line), 3$\sigma$ (solid red line) and 5$\sigma$ (solid blue line) are also shown \cite{Savage20091}.
The detection limit without dark matter hypothesis (section \ref{sssec:LC_LD}) is the solid black line; for $M_W>130$~GeV the modulation amplitude is negative in the [2,6] keV$_{\textnormal{ee}}$ energy interval, a result non considered in the one--tailed test because it is opposite to the DAMA/LIBRA signal.\label{fig:OneTailedAndLikelihood}}
\end{figure}

The detection limit under the dark matter hypothesis is shown in the Fig.~\ref{fig:OneTailedAndLikelihood}, taking the background shown in the Fig.~\ref{fig:background} and an exposure of $M\cdot{}T_M=112.5$ kg$\times 5$ years. In practically all the $3\sigma$ DAMA/LIBRA region, \mbox{ANAIS--112} can detect the annual modulation of the interaction rate of WIMPs with Na or I.
The region above the dashed black line of Fig.~\ref{fig:OneTailedAndLikelihood} is excluded because the dark matter rate in [2,6] keV$_{\textnormal{ee}}$ is greater than the observed one.

The one tailed $L_D$ of Eq.~(\ref{eq:aSingleEnergyBin}), deduced from the figure of merit Eq.~(\ref{eq:fom}), can be translated to the $(\sigma_{SI},M_W)$ plane, see the solid black line of the Fig.~\ref{fig:OneTailedAndLikelihood}. For ANAIS--112, it is numerically equivalent to the maximum likelihood ratio test under the dark matter hypothesis.

\section{ANAIS--112 in the [1,6] keV$_{\textnormal{ee}}$ energy interval} \label{sec:1to6}

We have focused on the [2,6] keV$_{\textnormal{ee}}$ energy interval, where DAMA/LIBRA has measured a positive modulation signal of dark matter. But ANAIS--112 is able to use more information from the background spectrum up to 1 keV$_\textnormal{ee}$.

\subsection{Model independent modulation}

The case of a single energy bin is not considered because $B(E)/\varepsilon(E)$ changes steeply below 2~keV$_\textnormal{ee}$ (Fig.~\ref{fig:background}).
To estimate $L_C$, a two--tailed test is carried out, $P(|\hat{b}|\leq L_C\mid b=0)=0.9$, since there is no modulation signal measured in this region allowing us to know its sign, $L_C=1.65~\sigma(\hat{b})$. $L_D$ is calculated as one--tailed probability, $P(|\hat{b}|>L_C\mid b=L_D)=0.9$, $L_D=L_C+1.28~\sigma(\hat{b})$, because of the choice $b=L_D$.

\subsubsection{Energy binning and segmented detector}

For $N=50$ ($\delta E=0.1$~keV$_\textnormal{ee}$), $\hat{b}_j$ and $\hat{b}_j^k$ are normal variables as in section~\ref{sssec:2to6_binned}.
Taking $var(\hat{b})$ of Eq.~(\ref{eq:bVarianceBinning}) and $\left\langle B/\varepsilon\right\rangle=24.6$~cpd/kg/keV$_\textnormal{ee}$ (Table~\ref{table:detectorsBkg[1,6]}), $L_D$ at 90\% C.L. (when $L_C$ is at 90\%) of ANAIS--112 after 5 years is:
\begin{equation}
L_D=(2.02\pm 0.06)\cdot 10^{-2} \textnormal{ cpd/kg/keV}_{\textnormal {ee}} \hspace{0.5cm}\left(90\%\textnormal{ C.L.}\right)
\label{eq:energyBinning[1,6]}
\end{equation}

\begin{figure}
	\centering
		\includegraphics[height=7.5cm]{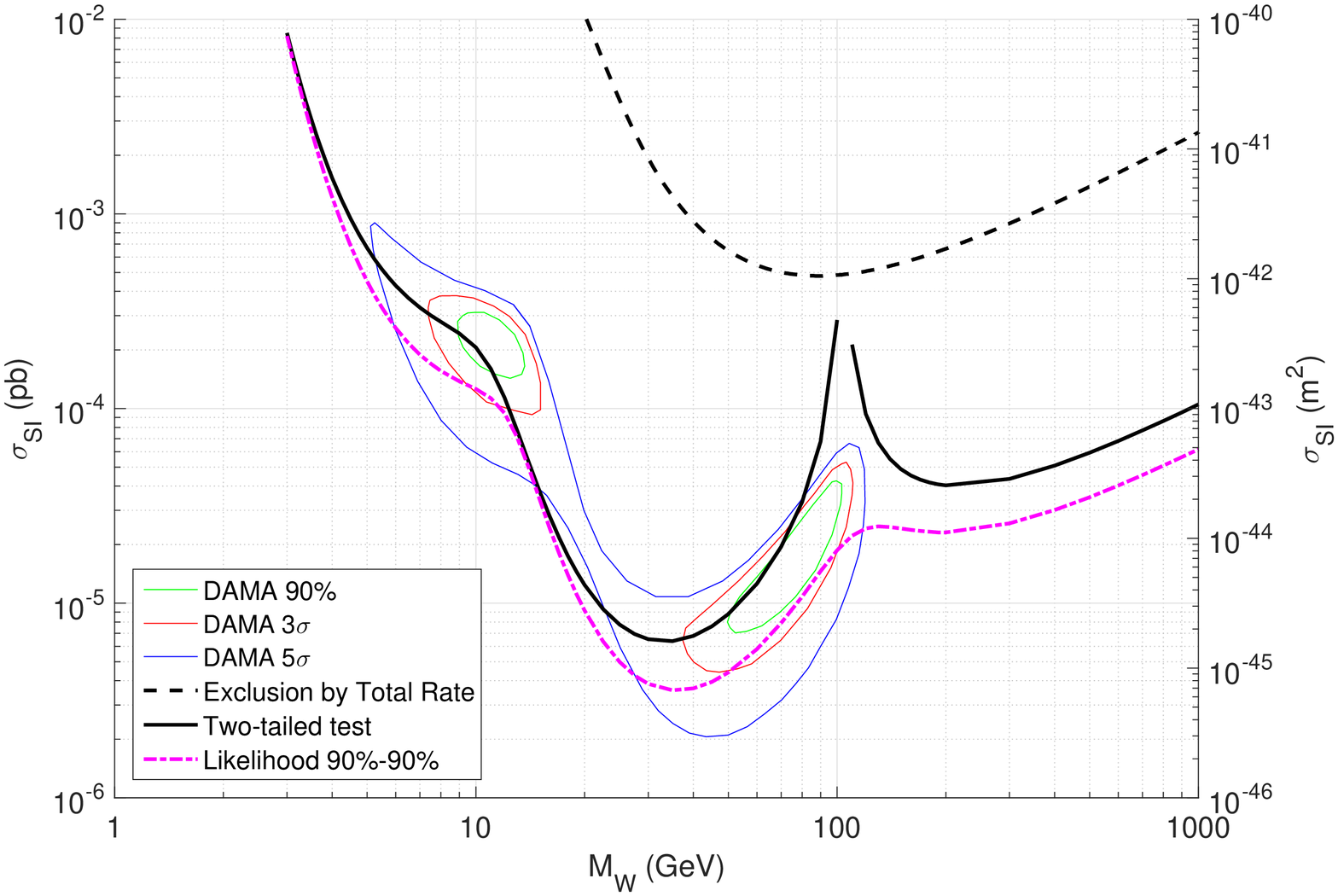}
	\caption{Result of the maximum likelihood test ratio for the detection limit in the [1,6] keV$_{\textnormal{ee}}$ window at 90\% C.L. (when critical limit is at 90\% C.L.) with 50 energy bins and segmented detector of ANAIS--112 after 5 years of measurement (dashed-dot magenta line). The exclusion by total rate for the spin-independent WIMP-nucleon cross section of ANAIS--112 is the dashed black line. DAMA/LIBRA regions at 90\% (solid green line), 3$\sigma$ (solid red line) and 5$\sigma$ (solid blue line) are also shown \cite{Savage20091}. The detection limit in the same conditions as before, but calculated as two--tailed test, is the solid black line; the discontinuity is due to the cancellation of the modulation amplitude in [1,6] keV$_{\textnormal{ee}}$ energy interval. \label{fig:TwoTailedAndLikelihood}}
\end{figure}

Taking each module separately, according to the Table \ref{table:detectorsBkg[1,6]} and the Eq.~(\ref{eq:9detectorsVariance}), $L_D=2.01\cdot{}10^{-2}$ cpd/kg/keV$_\textnormal{ee}$, very close to Eq.~(\ref{eq:energyBinning[1,6]}) because the nine values ${\left\langle B/\varepsilon\right\rangle}^k$ are close to ${\left\langle B/\varepsilon\right\rangle}$. Its plot in the $(\sigma_{SI},M_W)$ plane (Fig.~\ref{fig:TwoTailedAndLikelihood}) is worse than the one in [2,6] keV$_\textnormal{ee}$ (Fig.~\ref{fig:OneTailedAndLikelihood}) because it is deduced from a two--tailed test for $L_C$ and, without the dark matter hypothesis, there is no way of profiting the signal below 2~keV$_\textnormal{ee}$ to compensate the background increasing.

\subsection{Dark matter hypothesis}

The detection limit of ANAIS--112 at 90\% C.L. (for $L_C$ at 90\% C.L.) under the dark matter hypothesis is shown in the Fig.~\ref{fig:TwoTailedAndLikelihood}, taking the same exposure used in [2,6] keV$_\textnormal{ee}$ (Fig.~\ref{fig:OneTailedAndLikelihood}). The region of detection is now bigger for $M_W<50$~GeV, because the background increasing is compensated by a higher signal below 2~keV$_\textnormal{ee}$. It is better than the one obtained in the previous section because of the added condition $\hat{b}_j\approx b_j(\sigma,M_W)$, Eq.~(\ref{eq:chi2}).

\section{Conclusions}

We have estimated the detection limit at 90\% C.L., when the critical limit is at 90\% C.L., of ANAIS--112 for the annual modulation observed by DAMA/LIBRA. It is based on the measured background of the six modules D0 to D5. In the two considered scenarios (the [2,6] keV$_\textnormal{ee}$ used by DAMA/LIBRA and the [1,6] keV$_\textnormal{ee}$ also available for ANAIS--112), we conclude that after 5~years of measurement, ANAIS--112 can detect the annual modulation in the $3\sigma$ region compatible with the DAMA/LIBRA result.

We give a simple figure of merit that gives good estimates of $L_C$ and $L_D$ if the ratio $B(E)/\varepsilon(E)$ is smooth, as it is our case within [2,6] keV$_\textnormal{ee}$.

\section*{Acknowledgements}

This work has been supported by the Spanish Ministerio de Economía y Competitividad and the European Regional Development Fund (MINECO-FEDER) (FPA2011-23749 and FPA2014-55986-P), the Consolider-Ingenio 2010 Programme under grants MULTIDARK CSD2009-00064 and CPAN CSD2007-00042, and the Gobierno de Aragón and the European Social Fund (Group in Nuclear and Astroparticle Physics, GIFNA). P. Villar was supported by the MINECO Subprograma de Formación de Personal Investigador. We also acknowledge LSC and GIFNA staff for their support.


\bibliography{mybibfile}

\clearpage

\clearpage

\clearpage

\clearpage

\begin{table}
\begin{center}
\begin{tabular}{ccc} \hline \hline
 Module & $^{40}K$  & $^{210}Pb$ \\ \hline
 D0 &  1.4~$\pm$~0.2 & 3.15~$\pm$~0.10\\
 D1 &  1.1~$\pm$~0.2 & 3.15~$\pm$~0.10\\
 D2 &  1.1~$\pm$~0.2 & 0.7~$\pm$~0.1\\
 D3 &  0.60~$\pm$~0.06 & 1.8~$\pm$~0.1 \\
 D4 &  0.3~$\pm$~0.2  & 1.8~$\pm$~0.1 \\
 D5 &  0.8~$\pm$~0.2  & 0.70~$\pm$~0.05 \\
\hline \hline
\end{tabular}
\end{center}
\caption{Measured specific activity (mBq/kg) of $^{40}K$ and $^{210}Pb$ in the six modules D0 to D5. \label{table:40K-210Pb}}
\end{table}

\clearpage

\begin{landscape}

\begin{table}
\begin{center}
\begin{tabular}{cccccc} \hline \hline
 & $B$ & $B^\prime$ & $f(K)$ & $f(Pb)$ & $\left\langle B/\varepsilon\right\rangle$ \\
Module & \multicolumn{2}{c}{(cpd/kg/keV$_{\textnormal{ee}}$)} & \multicolumn{2}{c}{(cpd/keV$_{\textnormal{ee}}$/mBq)} & (cpd/kg/keV$_{\textnormal{ee}}$)\\ \hline
 D0 &  5.56 & 2.26$\pm$0.16 & 0.727 & 0.725 & 6.38 \\
 D1 &  5.52 & 2.44$\pm$0.16 & 0.727 & 0.725 & 6.30 \\
 D2 &  3.27 & 1.79$\pm$0.17 & 0.672 & 1.054 & 3.72 \\
 D3 &  4.10$\pm$0.42 & 2.08$\pm$0.28 & 0.700$\pm$0.028 & 0.890$\pm$0.165 & 4.67$\pm$0.47 \\
 D4 &  3.90$\pm$0.44 & 2.08$\pm$0.28 & 0.700$\pm$0.028 & 0.890$\pm$0.165 & 4.43$\pm$0.50 \\
 D5-D8 &  3.27$\pm$0.33 & 2.08$\pm$0.28 & 0.700$\pm$0.028 & 0.890$\pm$0.165 & 3.71$\pm$0.38 \\
 ANAIS--112 & 3.93$\pm$0.23 &  &  &  & 4.48$\pm$0.26 \\
\hline \hline
\end{tabular}
\end{center}
\caption{Measured background in the [2,6] keV$_{\textnormal{ee}}$ energy interval for D0, D1 and D2 modules and estimated background from D3 to D8 modules before ($2^{nd}$ column) and after ($6^{th}$ column) corrections for efficiencies have been applied. The estimated values for ANAIS--112 is listed in the last row. Columns $3^{rd}$ to $5^{th}$ list the values of $B^\prime$, $f(K)$ and $f(Pb)$ (see text). In the modules D0, D1 and D2 the statistical uncertainties are negligible for $B$ because of the long time measurements, and for $f(K)$ and $f(Pb)$ because they were estimated by Monte Carlo with enough precision.\label{table:detectorsBkg}}
\end{table}

\end{landscape}

\clearpage
\begin{landscape}

\begin{table}
\begin{center}
\begin{tabular}{cccccc} \hline \hline
 & $B$ & $B^\prime$ & $f(K)$ & $f(Pb)$ & $\left\langle B/\varepsilon\right\rangle$ \\
Module & \multicolumn{2}{c}{(cpd/kg/keV$_{\textnormal{ee}}$)} & \multicolumn{2}{c}{(cpd/keV$_{\textnormal{ee}}$/mBq)} & (cpd/kg/keV$_{\textnormal{ee}}$)\\ \hline
 D0 &  11.3 & 8.14$\pm$0.14 & 0.606 & 0.733 & 31.7 \\
 D1 &  9.88 & 6.90$\pm$0.14 & 0.606 & 0.733 & 26.1 \\
 D2 &  7.18 & 5.86$\pm$0.15 & 0.561 & 1.011 & 20.8 \\
 D3 &  8.66$\pm$0.88 & 6.74$\pm$0.83 & 0.584$\pm$0.023 & 0.872$\pm$0.139 & 24.8$\pm$2.5 \\
 D4 &  8.49$\pm$0.88 & 6.74$\pm$0.83 & 0.584$\pm$0.023 & 0.872$\pm$0.139 & 24.5$\pm$2.6 \\
 D5-D8 &  7.82$\pm$0.85 & 6.74$\pm$0.83 & 0.584$\pm$0.023 & 0.872$\pm$0.139 & 23.3$\pm$2.5 \\
 ANAIS--112 & 8.53$\pm$0.50 &  &  &  & 24.6$\pm$1.5 \\
\hline \hline
\end{tabular}
\end{center}
\caption{Same as Table \ref{table:detectorsBkg} in the [1,6] keV$_{\textnormal{ee}}$ energy interval. \label{table:detectorsBkg[1,6]}}
\end{table}

\end{landscape}

\end{document}